\documentclass[aps,prc,preprint,nofootinbib,amsmath,amssymb,floatfix]{revtex4-2}

\usepackage{graphicx}
\usepackage{dcolumn}
\usepackage{bm}
\usepackage{mathtools}
\usepackage{booktabs}
\usepackage[pdfpagelabels,pdfencoding=auto,psdextra]{hyperref}
\hypersetup{colorlinks=true,linkcolor=blue,citecolor=blue,urlcolor=blue}

\graphicspath{{./}{../results/}}

\begin{document}

\title{Structure and Binding of the \(pn\Omega^-\) System with the Physical-Point Lattice QCD \(N\Omega\) Interaction}
\author{Faisal Etminan}
 \email{fetminan@birjand.ac.ir}
\affiliation{
 Department of Physics, Faculty of Sciences, University of Birjand, Birjand 97175-615, Iran
}%
\affiliation{ Interdisciplinary Theoretical and Mathematical Sciences Program (iTHEMS), RIKEN, Wako 351-0198, Japan}
\author{Lucas Happ}
\email{lucas.happ@riken.jp}
\affiliation{Few-body Systems in Physics Laboratory, RIKEN Nishina Center, Wakō, Saitama 351-0198, Japan}
\affiliation{Institut für Quantenphysik and Center for Integrated Quantum Science and Technology (IQST), Universität Ulm, D-89069 Ulm, Germany}
\date{\today}%
\begin{abstract}
The binding and spatial structure of the \(pn\Omega^-\) system are investigated in the \(I(J^P)=0(5/2^+)\) channel using the physical-point HAL QCD \(N\Omega({}^{5}S_2)\) interaction. The pairwise Hamiltonian includes the spin-triplet \(pn\) interaction and the finite-size Coulomb attraction in the \(p\Omega^-\) pair. The three-body Schr\"odinger equation is solved with the Gaussian Expansion Method (GEM), including all three Jacobi rearrangement configurations. For the central HAL QCD potential at \(t/a=14\), the three-body state lies \(16.39\pm2.0\,(\mathrm{stat})\,{}^{+2.4}_{-2.2}\,(\mathrm{sys})\) MeV below the \(d+\Omega^-\) threshold, where the statistical uncertainty is obtained by mapping the lattice uncertainty of the \(n\Omega^-\) binding energy onto the three-body system and the systematic one from the Euclidean-time dependence of the potential. The three-body binding closely follows the \(n\Omega^-\) binding; earlier HAL QCD potentials with a larger \(N\Omega\) binding give a correspondingly deeper state. The spatial structure is examined through interparticle root-mean-square distances, pair and spectator radial distributions, and the correlation between the two Jacobi coordinates. Relative to the free deuteron and a frozen-deuteron folded reference, the full three-body solution exhibits a contracted \(pn\) pair and a substantially more compact spectator distribution.
\end{abstract}

\maketitle
\section{Introduction}
\label{sec:intro}

The \(\Omega^-\) baryon, with spin \(3/2\) and triply strange quark content, shares no valence quarks with a nucleon. The resulting absence of quark-exchange Pauli repulsion makes the \(N\Omega\) system a promising candidate for a strongly attractive baryon--baryon channel. The \(N\Omega\) interaction has been studied using constituent-quark models~\cite{PangPRC2004,HuangPRC922015}, effective field theories~\cite{Haidenbauer2017}, meson-exchange model~\cite{Sekihara2018} and the Pomeron exchange approaches~\cite{PomeronPRD2026}. Lattice QCD calculations by the HAL QCD Collaboration provide a first-principles determination of this interaction. Early calculations found strong attraction in the spin-2 \(S\)-wave channel, \({}^{5}S_2\)~\cite{etminan2014}, followed by a near-physical-point study that predicted a shallow state~\cite{Iritani2019prb}. More recently, a physical-point calculation with \((2+1)\)-flavor lattice QCD reported a shallow \(N\Omega\) pole near the unitary regime~\cite{zhang2026}, with binding energies of approximately \(0.61\) MeV for \(n\Omega\) and \(1.25\) MeV for \(p\Omega^-\), the latter including finite-size Coulomb effects.

The \(N\Omega\) interaction is also accessible through femtoscopy in relativistic heavy-ion collisions. Proton--\(\Omega^-\) correlations were proposed as a probe of the spin-2 \(N\Omega\) dibaryon~\cite{Morita2016} and measured by the STAR Collaboration in Au+Au collisions at \(\sqrt{s_{NN}}=200\) GeV~\cite{STAR2019}. The production of multistrange few-body systems, including \(\Omega NN\) and \(\Omega\Omega N\), has been investigated in coalescence-based studies~\cite{zhang2021production,EtminanPRC2025-111}.

The \(\Omega NN\) system is a natural extension of the \(N\Omega\) dibaryon problem. In the maximum-spin \(J^P=5/2^+\) channel, the nucleon pair can occupy the spin-triplet, isospin-singlet deuteron channel, while both \(N\Omega\) pairs interact in the attractive \({}^{5}S_2\) channel. Previous three-body studies have investigated this system using Faddeev methods and different \(NN\) and \(\Omega N\) interactions~\cite{Garcilazo2018,Garcilazo2019,ETMINAN2023122639,FilikhinPRC2025OmegaNuclei}, including the earlier HAL QCD \(N\Omega\) potentials~\cite{Garcilazo2019}.
The new physical-point interaction provides an opportunity to revisit the binding and spatial structure of this system with updated two-body input.

The present work investigates the \(pn\Omega^-\) system in the \(I(J^P)=0(5/2^+)\) channel using the physical-point HAL QCD \(N\Omega({}^{5}S_2)\) potential, a spin-triplet \(NN\) interaction, and the finite-size Coulomb interaction in the \(p\Omega^-\) pair. The three-body Schr\"odinger equation is solved using GEM with all three Jacobi rearrangement configurations. We determine the binding energy with statistical and systematic uncertainties derived from the published HAL QCD results, compare with earlier calculations based on the previous HAL QCD potentials, and analyze the structure of the state through interparticle distances, the \(pn\) and \(\Omega^--(pn)\) radial distributions, their coordinate correlation, and a comparison with the free deuteron and a frozen-deuteron folded \(d+\Omega^-\) reference.

\section{Model and method}
\label{sec:method}

\subsection{Hamiltonian and interactions}
\label{subsec:hamiltonian}

The \(pn\Omega^-\) system is considered in the isospin-spin-parity channel
$I(J^P)=0\left(\frac{5}{2}^{+}\right)$.
In this maximally aligned channel the nucleon pair is in the deuteron channel, $S_{NN}=1$, $I_{NN}=0$,
while each \(N\Omega\) pair interacts in the spin-2 channel, \({}^{5}S_2\), so that no spin recoupling between the pairs is required.
The system is described by a nonrelativistic pairwise Hamiltonian,
\begin{equation}
	H=T_{\mathrm{rel}}
	+V_{pn}^{(^3S_1)}
	+V_{p\Omega^-}^{(^5S_2)}
	+V_{n\Omega^-}^{(^5S_2)}
	+V_C^{p\Omega^-},
	\label{eq:hamiltonian}
\end{equation}
where \(T_{\mathrm{rel}}\) is the kinetic-energy operator after removal of the center-of-mass motion. The two strong \(N\Omega\) interactions are taken to be identical, corresponding to an isospin-symmetric approximation. The Coulomb interaction is included only in the \(p\Omega^-\) pair; it breaks exact isospin symmetry in the full Hamiltonian, and \(I=0\) labels the isospin channel of the strong-interaction problem.
For consistency with the physical-point HAL QCD calculation, the masses
$m_p=m_n=m_N=939.7~\mathrm{MeV}$ and $m_\Omega=1672.45~\mathrm{MeV}$ are adopted, together with $\hbar c=197.3269804~\mathrm{MeV\,fm}$ and $\alpha^{-1}=137.035999$.

For the proton--neutron pair, the Malfliet--Tjon two-range Yukawa potential in the spin-triplet ${}^3S_1$ channel is used~\cite{MALFLIET1969161},
\begin{equation}
	V_{NN}^{({}^3S_1)}(r)
	=
	-626.885\frac{e^{-1.55 r}}{r}
	+
	1438.72\frac{e^{-3.11 r}}{r},
	\label{eq:NN}
\end{equation}
with $r$ in fm and the potential in MeV. With the masses above it gives the deuteron binding energy $B_d=2.239~\mathrm{MeV}$, which determines the $d+\Omega^-$ threshold.

The strong nucleon--$\Omega$ interaction is taken from the
physical-point HAL QCD determination in the ${}^5S_2$ channel~\cite{zhang2026},
\begin{align}
	V_{N\Omega}(r)
	={}&
	a_1 e^{-b_1r^2}
	+a_2 e^{-b_2r^2}
	\nonumber\\
	&+
	a_3\left(1-e^{-(r/R)^2}\right)
	\left(\frac{e^{-m_\pi r}}{r}\right)^2,
	\label{eq:NOmega}
\end{align}
with $R=2.0~\mathrm{fm}$ and $m_\pi=137.1~\mathrm{MeV}$, converted to inverse fm using $m_\pi/(\hbar c)$. The parameters $a_1$, $a_2$, $b_1$, $b_2$, and $a_3$ are
obtained from fits to the HAL QCD potential at the Euclidean times $t/a=12$--$16$, with lattice spacing $a \simeq 0.084372$ fm; they are listed in Table~\ref{tab:potential}. The parametrization at $t/a=14$ is adopted as the central interaction, and the variation over $t/a=12$--$16$ is used as a systematic uncertainty. The different Euclidean-time slices are not statistically independent realizations of the potential.
The HAL QCD analysis reports the two-body binding energies
\begin{align}
	B_{n\Omega}
	&=
	0.61^{+0.41}_{-0.30}
	{}^{+0.60}_{-0.18}
	~\mathrm{MeV},
	\label{eq:nOmega_binding_HAL}\\
	B_{p\Omega}
	&=
	1.25^{+0.46}_{-0.39}
	{}^{+0.69}_{-0.25}
	~\mathrm{MeV},
	\label{eq:pOmega_binding_HAL}
\end{align}
where the first uncertainty is statistical and the second systematic, and the $p\Omega^-$ result includes the finite-size Coulomb
interaction. The fitted $t/a=14$ potential gives $B_{n\Omega}=0.622$ and $B_{p\Omega}=1.263~\mathrm{MeV}$ in the present implementation, in agreement with these values.
The same strong parametrization is employed for the $p\Omega^-$ and $n\Omega^-$ pairs.

\begin{table*}[t]
	\centering
	\caption{Parameters of the physical-point HAL QCD
		$N\Omega$ potential at different Euclidean-time
		separations~\cite{zhang2026}. The uncertainties in parentheses are the statistical uncertainties of the individual fit parameters. The coefficients $a_1$ and $a_2$ are
		given in MeV, $b_1$ and $b_2$ in $\mathrm{fm}^{-2}$,
		and $a_3$ in $\mathrm{MeV\,fm^2}$.}
	\label{tab:potential}
	\setlength{\tabcolsep}{9pt}
	\begin{tabular}{c r r r r r}
		\toprule
		$t/a$
		& $a_1$
		& $b_1$
		& $a_2$
		& $b_2$
		& $a_3$ \\
		\midrule
		12 &
		$-312.7(2.2)$ &
		$51.8(1.1)$ &
		$-76.3(7.5)$ &
		$1.44(4)$ &
		$-289.0(46.9)$ \\

		13 &
		$-314.8(3.9)$ &
		$50.6(2.3)$ &
		$-83.9(11.3)$ &
		$1.43(8)$ &
		$-244.1(78.1)$ \\

		14 &
		$-311.5(6.7)$ &
		$49.0(2.4)$ &
		$-94.7(16.6)$ &
		$1.38(10)$ &
		$-162.6(111.1)$ \\

		15 &
		$-307.1(15.6)$ &
		$52.6(4.4)$ &
		$-76.2(17.2)$ &
		$1.47(19)$ &
		$-265.3(118.3)$ \\

		16 &
		$-285.4(24.1)$ &
		$56.4(8.9)$ &
		$-56.3(17.3)$ &
		$1.73(46)$ &
		$-358.8(131.5)$ \\
		\bottomrule
	\end{tabular}
\end{table*}

The Coulomb interaction between the proton and the $\Omega^-$ is
\begin{equation}
	V_C^{p\Omega^-}(r)
	=
	-\frac{\alpha\hbar c}{r}F_C(r),
	\label{eq:Coulomb}
\end{equation}
where $F_C(r)$ accounts for the finite charge distributions
of the proton and the $\Omega^-$~\cite{zhang2026}, with the charge radii
$r_p=0.841~\mathrm{fm}$~\cite{RevModPhys.97.025002} and $r_\Omega=0.571(18)~\mathrm{fm}$~\cite{PhysRevD.92.114515}. Both charge distributions are described by dipole form factors~\cite{LyuPRL2021},
$F(k)=\Lambda^4/(k^2+\Lambda^2)^2$, corresponding to the normalized densities $\rho(r)=\Lambda^3e^{-\Lambda r}/(8\pi)$ with RMS charge radius $r_{\mathrm{ch}}=\sqrt{12}/\Lambda$. With
$\Lambda_p={\sqrt{12}}/{r_p}$, $\Lambda_\Omega={\sqrt{12}}/{r_\Omega}$, $a=\Lambda_p^2$, and $b=\Lambda_\Omega^2$,
the finite-size factor in Eq.~(\ref{eq:Coulomb}) for the convolution of
the two charge distributions is~\cite{zhang2026}
\begin{align}
	F_C(r)
	={}&
	\Lambda_p^4\Lambda_\Omega^4
	\Bigg[
	-\frac{2(1-e^{-\Lambda_p r})}
	{a(b-a)^3}
	+\frac{2(1-e^{-\Lambda_\Omega r})}
	{b(b-a)^3}
	\nonumber\\
	&+
	\frac{
		2(1-e^{-\Lambda_p r})
		-\Lambda_p r e^{-\Lambda_p r}
	}{
		2a^2(b-a)^2
	}
	+
	\frac{
		2(1-e^{-\Lambda_\Omega r})
		-\Lambda_\Omega r e^{-\Lambda_\Omega r}
	}{
		2b^2(b-a)^2
	}
	\Bigg].
	\label{eq:FC}
\end{align}
It approaches unity at large separation and regularizes the short-distance behavior, \(V_C^{p\Omega^-}(0)=-2.19\)~MeV. This closed form agrees with a direct numerical folding of the two form factors in momentum space to better than \(10^{-11}\).

\subsection{Gaussian Expansion Method}
\label{subsec:GEM}

The three-body Schr\"odinger equation is solved using the Gaussian Expansion Method (GEM)~\cite{HIYAMA2003223,etminan2026nnomegaccc}, a variational approach. Three Jacobi coordinate sets are used, \(c=1,2,3\), in which the \(n\Omega^-\), \(p\Omega^-\), and \(pn\) pair, respectively, is described by the relative coordinate \(\boldsymbol r_c\), and the remaining particle by its coordinate \(\boldsymbol R_c\) relative to the center of mass of the pair. In the \(pn\) set,
\begin{equation}
	\boldsymbol r=\boldsymbol r_p-\boldsymbol r_n,
	\qquad
	\boldsymbol R=
	\boldsymbol r_{\Omega^-}
	-\frac{\boldsymbol r_p+\boldsymbol r_n}{2},
	\label{eq:Omega_deuteron_coordinate}
\end{equation}
The coordinate \(\boldsymbol R\) is defined with respect to the center of mass of the two-nucleon subsystem; it does not imply that the \(pn\) pair is assumed to remain an isolated deuteron. The wave function is expanded as
\begin{equation}
	\Psi_{JM}=
	\sum_{c=1}^{3}
	\sum_{n=1}^{n_{\max}}
	\sum_{N=1}^{N_{\max}}
	C_{nNc}\,
	\phi_n^{(c)}(r_c)
	\psi_N^{(c)}(R_c)\,
	\chi_{JM}^{(c)},
	\label{eq:wavefunction_expansion}
\end{equation}
where \(C_{nNc}\) are variational coefficients and \(\chi_{JM}^{(c)}\) are the spin functions. The radial basis functions are
\begin{equation}
	\phi_n^{(c)}(r_c)
	=\mathcal{N}_n
	e^{-(r_c/r_n)^2},
	\qquad
	\psi_N^{(c)}(R_c)
	=\mathcal{N}_N
	e^{-(R_c/R_N)^2},
	\label{eq:Gaussian_basis}
\end{equation}
with normalization constants \(\mathcal{N}_n\) and \(\mathcal{N}_N\) and ranges in geometric progression,
$r_n=r_{\min}(r_{\max}/r_{\min})^{(n-1)/(n_{\max}-1)}$ and $R_N=R_{\min}(R_{\max}/R_{\min})^{(N-1)/(N_{\max}-1)}$. The basis functions are \(S\) waves (\(l=L=0\)); the local potentials are not projected onto \(S\) waves, so that the \(S\)-wave functions of one set also contribute higher partial waves in the others. The resulting generalized eigenvalue problem is solved with the FewBodyToolkit.jl package~\cite{Happ2026a,Happ2026b}, and the spatial observables are evaluated with the normalized ground-state wave function.

We use $r_{\min}=0.01$~fm, $R_{\min}=0.001$~fm, and $r_{\max}=R_{\max}=50$~fm, with $n_{\max}=N_{\max}=20$ for energies and root-mean-square (RMS) distances and $28$ for the structure observables; the thresholds are computed in the same basis. 
The separation energy at $t/a=14$ changes by $1.0$~keV from $n_{\max}=N_{\max}=20$ to $24$ and by $0.05$~keV from $24$ to $28$, and the RMS distances change by less than \(0.001\)~fm.

\subsection{Observables}
\label{subsec:observables}

Besides the RMS interparticle distances $r_{ij}^{\mathrm{rms}}=\sqrt{\langle(\boldsymbol r_i-\boldsymbol r_j)^2\rangle}$ and $R_{\mathrm{rms}}=\sqrt{\langle R^2\rangle}$, radial probability distributions are calculated for the two coordinates of Eq.~(\ref{eq:Omega_deuteron_coordinate}),
\begin{align}
	P_{pn}(r)
	&=
	r^2
	\int d\Omega_r
	\int d^3R\,
	\sum_{\sigma}
	\left|
	\Psi_{JM}(\boldsymbol r,\boldsymbol R;\sigma)
	\right|^2,
	\label{eq:pn_radial_probability}
	\\
	P_{\Omega-(pn)}(R)
	&=
	R^2
	\int d\Omega_R
	\int d^3r\,
	\sum_{\sigma}
	\left|
	\Psi_{JM}(\boldsymbol r,\boldsymbol R;\sigma)
	\right|^2,
	\label{eq:spectator_radial_probability}
\end{align}
where \(\sigma\) denotes the spin projections. Both are normalized to unity, and the cumulative probabilities are
\begin{equation}
	\mathcal P_{pn}(r<r_0)
	=
	\int_0^{r_0}P_{pn}(r)\,dr, \qquad
	\mathcal P_{\Omega-(pn)}(R<R_0)
	=
	\int_0^{R_0}P_{\Omega-(pn)}(R)\,dR.
	\label{eq:cumulative}
\end{equation}
The correlation between the two coordinates is characterized by
\begin{equation}
	\rho_{r^{2},R^{2}}
	=
	\frac{
		\langle r_{pn}^{2}R^{2}\rangle
		-
		\langle r_{pn}^{2}\rangle
		\langle R^{2}\rangle
	}{
		\sqrt{
			\left[\langle r_{pn}^{4}\rangle-\langle r_{pn}^{2}\rangle^{2}\right]
			\left[\langle R^{4}\rangle-\langle R^{2}\rangle^{2}\right]
		}
	}.
	\label{eq:coordinate_correlation}
\end{equation}
All moments and cumulative probabilities are evaluated directly from the GEM wave function.

As references for the structure, we use the free deuteron obtained with the same \(NN\) interaction and a frozen-deuteron folded \(d+\Omega^-\) model, in which the deuteron is kept in its isolated state and the \(d\Omega^-\) interaction is obtained by folding the \(N\Omega\) and Coulomb interactions over the radial deuteron density \(\rho_d(r)\), normalized as \(\int_0^\infty\rho_d(r)\,dr=1\),
\begin{equation}
	V_{d\Omega}(R)=\int_0^\infty dr\,\rho_d(r)\,\frac{1}{2}\int_{-1}^{1}d\mu\,
	\left[2V_{N\Omega}(s)+V_C^{p\Omega^-}(s)\right],
	\quad s=\sqrt{R^2+r^2/4-rR\mu},
	\label{eq:folding}
\end{equation}
with the reduced mass $2m_Nm_\Omega/(2m_N+m_\Omega)$. 

\section{Results and Discussion}
\label{sec:results}

\subsection{Binding energy}
\label{subsec:binding}

We first determine the energy of the three-body ground state for the five HAL QCD parametrizations. Table~\ref{tab:binding} lists the two-body binding energies, the three-body binding energy relative to the three-particle threshold, \(B_3=-E_3\), and the separation energy relative to the \(d+\Omega^-\) threshold,
\begin{equation}
	B_{\mathrm{sep}}=E_{d+\Omega}-E_3=B_3-B_d,
\end{equation}
together with the RMS distances. Among the two-body subsystems, the deuteron provides the lowest two-body-plus-spectator threshold, and for all \(t/a\) the three-body state lies far below it. For the central interaction at \(t/a=14\), \(B_3=18.634\)~MeV and \(B_{\mathrm{sep}}=16.394\)~MeV. The separation energy is appreciably larger than the binding energy of any two-body subsystem.

\begin{table*}[t]
	\centering
	\caption{Two-body binding energies \(B_{n\Omega}\) and \(B_{p\Omega}\) (with Coulomb), three-body binding energy \(B_3\) relative to the three-particle threshold, separation energy \(B_{\mathrm{sep}}\) relative to the \(d+\Omega^-\) threshold (\(B_d=2.239\) MeV), and RMS distances, for the HAL QCD potentials at Euclidean times \(t/a=12\)--\(16\), with Coulomb interaction. Energies in MeV, distances in fm.}
	\label{tab:binding}
	\setlength{\tabcolsep}{7pt}
	\begin{tabular}{ccccccccc}
		\toprule
		$t/a$ & $B_{n\Omega}$ & $B_{p\Omega}$ & $B_3$ & $B_{\mathrm{sep}}$
		& $r_{pn}^{\mathrm{rms}}$
		& $r_{p\Omega}^{\mathrm{rms}}$
		& $r_{n\Omega}^{\mathrm{rms}}$
		& $R_{\mathrm{rms}}$ \\
		\midrule
		12 & 1.238 & 1.965 & 21.03 & 18.79 & 2.004 & 1.750 & 1.763 & 1.443 \\
		13 & 1.025 & 1.729 & 20.29 & 18.05 & 2.011 & 1.759 & 1.773 & 1.452 \\
		14 & 0.622 & 1.263 & 18.63 & 16.39 & 2.036 & 1.791 & 1.805 & 1.482 \\
		15 & 0.688 & 1.330 & 18.49 & 16.25 & 2.059 & 1.820 & 1.834 & 1.510 \\
		16 & 0.427 & 0.992 & 16.43 & 14.20 & 2.134 & 1.921 & 1.935 & 1.606 \\
		\bottomrule
	\end{tabular}
\end{table*}

The contribution of the electromagnetic interaction is obtained by switching off the finite-size Coulomb potential while keeping all strong interactions, masses, and numerical parameters unchanged (Table~\ref{tab:coulomb}). The Coulomb attraction increases the binding by
$\Delta B_C=B_3^{(C)}-B_3^{(0)}=1.060$~MeV at $t/a=14$, about \(6\%\) of the Coulomb-free binding energy, and by \(1.0\) to \(1.1\)~MeV for all \(t/a\). The expectation value of the Coulomb potential in the ground state, $\langle V_C\rangle=-1.064~\mathrm{MeV}$, is close to $\Delta B_C$; the small difference reflects the rearrangement of the wave function induced by the Coulomb interaction.

\begin{table}[t]
	\centering
	\caption{Three-body binding energy \(B_3\), separation energy \(B_{\mathrm{sep}}\) (MeV), and RMS distances (fm) at \(t/a=14\), with and without the Coulomb interaction.}
	\label{tab:coulomb}
	\setlength{\tabcolsep}{12pt}
	\begin{tabular}{lcc}
		\toprule
		& with Coulomb & without Coulomb \\
		\midrule
		$B_3$ & 18.634 & 17.574 \\
		$B_{\mathrm{sep}}$ & 16.394 & 15.334 \\
		$r_{pn}^{\mathrm{rms}}$ & 2.036 & 2.050 \\
		$r_{p\Omega}^{\mathrm{rms}}$ & 1.791 & 1.816 \\
		$r_{n\Omega}^{\mathrm{rms}}$ & 1.805 & 1.816 \\
		$R_{\mathrm{rms}}$ & 1.482 & 1.499 \\
		\bottomrule
	\end{tabular}
\end{table}

The separation energy varies between \(14.20\) and \(18.79\)~MeV over \(t/a=12\)--\(16\) (Table~\ref{tab:binding}). Following the HAL QCD analysis, we take this spread as the systematic uncertainty, \({}^{+2.4}_{-2.2}\)~MeV relative to the central value.

\begin{figure}[t]
	\includegraphics[width=\columnwidth]{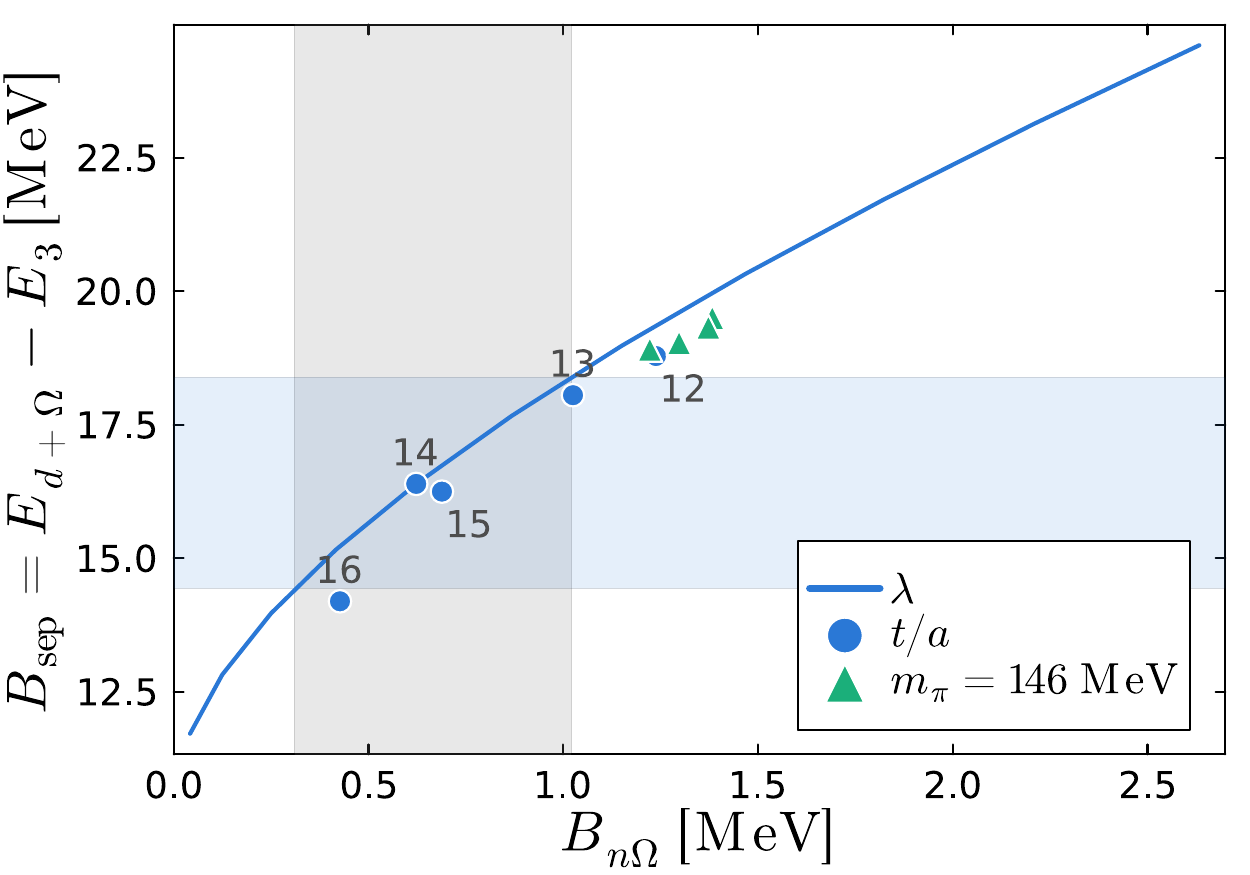}
	\caption{Separation energy \(B_{\mathrm{sep}}\) of the \(pn\Omega^-\) state relative to the \(d+\Omega^-\) threshold versus the \(n\Omega^-\) binding energy, with Coulomb. Line: central potential (\(t/a=14\)) scaled by \(\lambda=0.90\) to \(1.15\). Circles: HAL QCD potentials at \(t/a=12\)--\(16\) (labels). Triangles: earlier HAL QCD potential at \(m_\pi=146\)~MeV~\cite{Iritani2019prb}, parameter sets P1--P4 as given in Ref.~\cite{Garcilazo2019}, calculated here with the same Coulomb interaction and masses. Vertical band: statistical uncertainty of \(B_{n\Omega}\) from Ref.~\cite{zhang2026}; horizontal band: resulting range of \(B_{\mathrm{sep}}\).}
	\label{fig:b3_vs_bno}
\end{figure}

To translate the statistical uncertainty of the lattice potential into the three-body system, we use the relation between the three-body and the two-body binding shown in Fig.~\ref{fig:b3_vs_bno}. The line is obtained by scaling the central potential, \(V_{N\Omega}\to\lambda V_{N\Omega}\) with \(\lambda=0.90\) to \(1.15\), so that only its strength changes; for each \(\lambda\), \(B_{n\Omega}\) and \(B_{\mathrm{sep}}\) are calculated with the same numerical setup. The points show the HAL QCD potentials at \(t/a=12\)--\(16\), which differ in both strength and shape. The points lie close to the line: the three-body binding follows mainly the \(n\Omega^-\) binding energy, while the shape of the potential changes it only moderately at fixed \(B_{n\Omega}\). Assuming that the statistical fluctuations of the potential likewise act mainly on its strength, we determine the values \(\lambda=0.960\) and \(1.039\) for which \(B_{n\Omega}\) reaches the edges of the statistical band of Eq.~(\ref{eq:nOmega_binding_HAL}), \(0.31\) and \(1.02\)~MeV. The corresponding three-body calculations give \(B_{\mathrm{sep}}=14.43\) and \(18.39\)~MeV. This procedure uses only the published HAL QCD results and does not require the covariance of the fit parameters, which is not available. 
The final result for the separation energy is
\begin{equation}
	B_{\mathrm{sep}}=16.39\pm2.0\,(\mathrm{stat})\,{}^{+2.4}_{-2.2}\,(\mathrm{sys})~\mathrm{MeV}.
	\label{eq:final}
\end{equation}

These results are predictions within the adopted interaction model. Beyond the quoted uncertainties, they depend on the treatment of the \(NN\) interaction, the partial-wave space, and the absence of three-body forces and coupled inelastic channels.
Although the present calculation does not include explicit coupling to the inelastic \(\Lambda\Xi N\) and \(\Sigma\Xi N\) channels, the \(J^P=5/2^+\) quantum numbers impose an important restriction on these decay modes. Since all three baryons in either final state have spin \(1/2\), their total spin cannot exceed \(3/2\). Consequently, the final states cannot accommodate \(J=5/2\) when all relative orbital angular momenta are zero. Decays into these channels therefore require nonzero orbital angular momenta and are expected to be suppressed by the associated centrifugal barriers. This suggests that the \(pn\Omega^-\) state may be relatively narrow, although a quantitative estimate of its width requires an explicit coupled-channel calculation.

\subsection{Comparison with earlier calculations}
\label{subsec:comparison}

We now relate the present result to the earlier calculation of Garcilazo and Valcarce~\cite{Garcilazo2019}, who solved the Faddeev equations for the same channel with the same \(NN\) interaction, \(S\)-wave two-body \(t\)-matrices, and the earlier HAL QCD \(N\Omega\) potential at \(m_\pi=146\)~MeV~\cite{Iritani2019prb} with four sets of fit parameters, denoted P1--P4; the Coulomb interaction was added in a separate step as a screened point-charge potential. To compare the methods directly, we repeat their calculation with GEM, using the same potentials and the masses \(m_N=938.9\) and \(m_\Omega=1672.45\)~MeV. The results are listed in Table~\ref{tab:gv}. The two-body binding energies agree. The three-body binding energies obtained here are larger by \(0.4\) to \(0.7\)~MeV, about \(3\%\), consistently for all four potentials; the same offset is found with the heavier masses of the HAL QCD calculation. A possible origin is the restriction of the Faddeev calculation to configurations with all three particles in \(S\)-wave states~\cite{Garcilazo2019}, whereas in the present approach the local potentials also act in higher partial waves (Sec.~\ref{subsec:GEM}). With the screened point-Coulomb potential of Ref.~\cite{Garcilazo2019}, the present calculation gives an additional binding of \(1.3\)~MeV for all four potentials, as found there, and the same offset in the total binding energies. The finite-size Coulomb interaction of the present work gives a smaller increase of about \(1.08\)~MeV.

\begin{table}[t]
	\centering
	\caption{Binding energies (MeV) for the earlier HAL QCD potential~\cite{Iritani2019prb} with the parameter sets P1--P4 of Ref.~\cite{Garcilazo2019}, without Coulomb interaction and with \(m_N=938.9\) MeV: \(N\Omega\) binding energy and three-body binding energy relative to the three-particle threshold, from the present GEM calculation and from the Faddeev calculation of Ref.~\cite{Garcilazo2019}.}
	\label{tab:gv}
	\setlength{\tabcolsep}{9pt}
	\begin{tabular}{lcccc}
		\toprule
		& \multicolumn{2}{c}{$B_{N\Omega}$} & \multicolumn{2}{c}{$B_3$} \\
		\cmidrule{2-3}\cmidrule{4-5}
		Potential & GEM & Ref.~\cite{Garcilazo2019} & GEM & Ref.~\cite{Garcilazo2019} \\
		\midrule
		P1 & 1.291 & 1.29 & 20.16 & 19.6 \\
		P2 & 1.376 & 1.38 & 20.62 & 20.0 \\
		P3 & 1.216 & 1.29\footnote{Value as given in Ref.~\cite{Garcilazo2019}, identical to P1.} & 20.04 & 19.6 \\
		P4 & 1.366 & 1.37 & 20.45 & 19.9 \\
		\bottomrule
	\end{tabular}
\end{table}

With the Coulomb interaction and the masses of the present work, the earlier HAL QCD potential gives \(B_{\mathrm{sep}}=18.9\) to \(19.5\)~MeV for the four parameter sets, shown as triangles in Fig.~\ref{fig:b3_vs_bno}. They follow the same trend as the physical-point potentials: their larger \(N\Omega\) binding energies, \(1.2\) to \(1.4\)~MeV, give a correspondingly deeper three-body state, close to the \(t/a=12\) result. The physical-point interaction, with its weaker \(n\Omega^-\) binding, therefore leads to a shallower, although still deeply bound, \(pn\Omega^-\) state.

\subsection{Spatial structure}
\label{subsec:structure}

\begin{figure*}[t]
	\includegraphics[width=\textwidth]{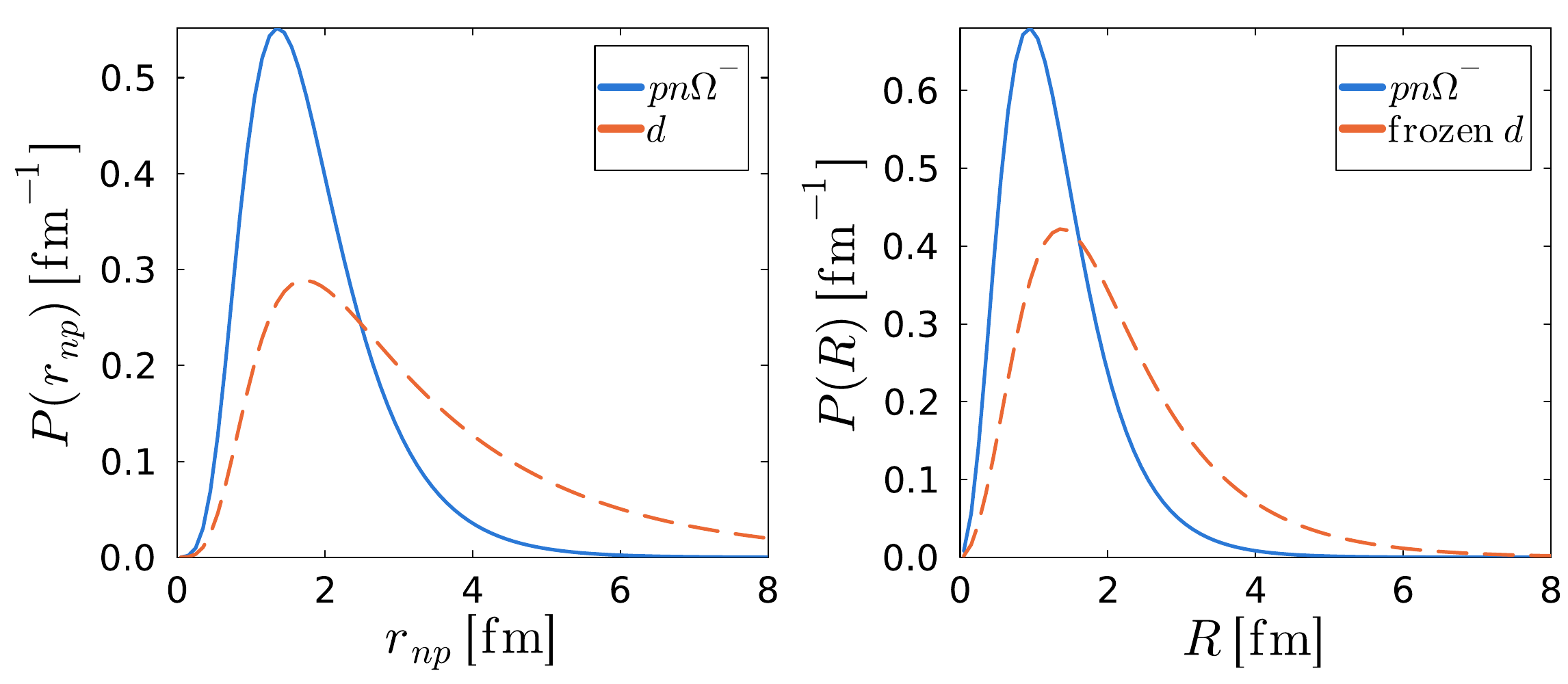}
	\caption{Radial probability distributions at \(t/a=14\) with Coulomb. Left: \(P_{pn}(r)\), Eq.~(\ref{eq:pn_radial_probability}), in the \(pn\Omega^-\) state (solid) and in the free deuteron (dashed). Right: \(P_{\Omega-(pn)}(R)\), Eq.~(\ref{eq:spectator_radial_probability}), in the \(pn\Omega^-\) state (solid) and in the frozen-deuteron folded model, Eq.~(\ref{eq:folding}) (dashed). The three-body distributions are averages over bins of \(0.1\) fm. All distributions are normalized to unity.}
	\label{fig:structure}
\end{figure*}

We next characterize the spatial structure of the state, using the RMS distances in Table~\ref{tab:binding} and the radial distributions of Sec.~\ref{subsec:observables}, shown in Fig.~\ref{fig:structure} and Table~\ref{tab:cumulative}. The proton--neutron separation is somewhat larger than either nucleon--$\Omega$ separation, although all three distances are on a comparable length scale, and the \(\Omega^-\) is located close to the \(pn\) center of mass, \(R_{\mathrm{rms}}=1.482\)~fm. The ground state is therefore spatially compact. The RMS distances satisfy the identity \(\langle r_{p\Omega}^2\rangle+\langle r_{n\Omega}^2\rangle=2\langle R^2\rangle+\frac{1}{2}\langle r_{pn}^2\rangle\), a consistency check of the Jacobi-coordinate evaluation. Over \(t/a=12\)--\(16\), \(r_{pn}^{\mathrm{rms}}\) varies from \(2.00\) to \(2.13\)~fm and \(R_{\mathrm{rms}}\) from \(1.44\) to \(1.61\)~fm, so that all potentials give a compact configuration.

Without the Coulomb interaction, the proton and neutron are equivalent in the present model, with identical nucleon--\(\Omega^-\) RMS distances of \(1.816\)~fm (Table~\ref{tab:coulomb}). The Coulomb interaction contracts the proton--\(\Omega^-\) distance most, to \(1.791\)~fm, but also reduces \(r_{n\Omega}^{\mathrm{rms}}\), \(r_{pn}^{\mathrm{rms}}\), and \(R_{\mathrm{rms}}\) slightly: although it acts only between the proton and the \(\Omega^-\), it affects the overall geometry through the correlated wave function.

\begin{table}[t]
	\centering
	\caption{Cumulative probabilities, Eq.~(\ref{eq:cumulative}), and RMS values of the \(pn\) and spectator coordinates at \(t/a=14\) with Coulomb, for the \(pn\Omega^-\) state, the free deuteron, and the frozen-deuteron folded \(d+\Omega^-\) model.}
	\label{tab:cumulative}
	\setlength{\tabcolsep}{9pt}
	\begin{tabular}{lcccc}
		\toprule
		& \multicolumn{2}{c}{$\mathcal P_{pn}(r<x)$} & \multicolumn{2}{c}{$\mathcal P_{\Omega-(pn)}(R<x)$} \\
		\cmidrule{2-3}\cmidrule{4-5}
		$x$ (fm) & \(pn\Omega^-\) & free \(d\) & \(pn\Omega^-\) & folded \(d+\Omega^-\) \\
		\midrule
		1 & 0.150 & 0.058 & 0.389 & 0.164 \\
		2 & 0.652 & 0.323 & 0.854 & 0.561 \\
		3 & 0.900 & 0.564 & 0.973 & 0.810 \\
		4 & 0.974 & 0.725 & 0.995 & 0.922 \\
		\midrule
		RMS (fm) & 2.036 & 3.972 & 1.482 & 2.441 \\
		\bottomrule
	\end{tabular}
\end{table}

The \(pn\) pair is substantially more compact in the three-body state than in the free deuteron: its RMS separation is about half the deuteron value, and the probability within \(2\)~fm is doubled (Table~\ref{tab:cumulative}). The \(pn\) subsystem is thus strongly modified by the \(\Omega^-\) and cannot be represented by an unchanged deuteron wave function. Consistently, the frozen-deuteron folded model gives a much weaker binding, \(B_{d\Omega}^{\mathrm{folded}}=4.76\)~MeV, and a spectator distribution that extends to larger distances; the three-body spectator RMS distance is about \(40\%\) smaller. 
The folded model is therefore used only as a structural reference, not as an equivalent three-body calculation.

The two coordinates are moderately correlated. With \(\langle r_{pn}^2\rangle=4.145~\mathrm{fm}^2\), \(\langle R^2\rangle=2.196~\mathrm{fm}^2\), \(\langle r_{pn}^4\rangle=37.20~\mathrm{fm}^4\), \(\langle R^4\rangle=11.83~\mathrm{fm}^4\), and \(\langle r_{pn}^2R^2\rangle=12.15~\mathrm{fm}^4\), Eq.~(\ref{eq:coordinate_correlation}) gives \(\rho_{r^{2},R^{2}}=0.257\): configurations with larger \(pn\) separation tend, on average, to have a larger spectator distance.

\section{Summary and conclusions}
\label{sec:conclusion}

The binding and spatial structure of the \(pn\Omega^-\) system in the \(I(J^P)=0(5/2^+)\) channel have been investigated using the physical-point HAL QCD \(N\Omega({}^{5}S_2)\) interaction. The three-body Schr\"odinger equation was solved with GEM, including all three Jacobi rearrangement configurations, the spin-triplet \(pn\) interaction, and the finite-size Coulomb attraction in the \(p\Omega^-\) pair.

For the central HAL QCD parametrization at \(t/a=14\), the state lies \(16.39\pm2.0\,(\mathrm{stat})\,{}^{+2.4}_{-2.2}\,(\mathrm{sys})\) MeV below the \(d+\Omega^-\) threshold, corresponding to a three-body binding energy of \(18.63\) MeV. The Coulomb interaction contributes about \(1.06\) MeV. The statistical uncertainty follows from the close relation between the three-body and the \(n\Omega^-\) binding energies, which maps the published statistical uncertainty of \(B_{n\Omega}\) onto the three-body system; the systematic uncertainty is the spread over the Euclidean times \(t/a=12\)--\(16\). The earlier HAL QCD potentials, which have larger \(N\Omega\) binding energies, give a correspondingly deeper state, and a GEM calculation with these potentials agrees with the earlier Faddeev result to about \(3\%\).

The spatial observables reveal a substantial rearrangement of both Jacobi coordinates. The \(pn\) RMS separation decreases from \(3.972\) fm in the free deuteron to \(2.036\) fm in the three-body state, and the spectator RMS distance is \(1.482\) fm, compared with \(2.441\) fm in the frozen-deuteron folded reference. An unchanged deuteron is therefore not an adequate description of the calculated three-body state. The pair and spectator coordinates are moderately correlated, \(\rho_{r^{2},R^{2}}=0.257\).

These results support a compact \(pn\Omega^-\) state within the adopted pairwise interaction model. A propagation of the lattice-potential uncertainties based on correlated samples and an assessment of higher partial waves of the two-body interactions would be valuable extensions. The updated physical-point interaction and the structural information reported here may provide useful input for future studies of multistrange few-body systems and their possible femtoscopic signatures in heavy-ion collisions.


\bibliography{Refs}

\begin{thebibliography}{24}%
\makeatletter
\providecommand \@ifxundefined [1]{%
 \@ifx{#1\undefined}
}%
\providecommand \@ifnum [1]{%
 \ifnum #1\expandafter \@firstoftwo
 \else \expandafter \@secondoftwo
 \fi
}%
\providecommand \@ifx [1]{%
 \ifx #1\expandafter \@firstoftwo
 \else \expandafter \@secondoftwo
 \fi
}%
\providecommand \natexlab [1]{#1}%
\providecommand \enquote  [1]{``#1''}%
\providecommand \bibnamefont  [1]{#1}%
\providecommand \bibfnamefont [1]{#1}%
\providecommand \citenamefont [1]{#1}%
\providecommand \href@noop [0]{\@secondoftwo}%
\providecommand \href [0]{\begingroup \@sanitize@url \@href}%
\providecommand \@href[1]{\@@startlink{#1}\@@href}%
\providecommand \@@href[1]{\endgroup#1\@@endlink}%
\providecommand \@sanitize@url [0]{\catcode `\\12\catcode `\$12\catcode
  `\&12\catcode `\#12\catcode `\^12\catcode `\_12\catcode `\%12\relax}%
\providecommand \@@startlink[1]{}%
\providecommand \@@endlink[0]{}%
\providecommand \url  [0]{\begingroup\@sanitize@url \@url }%
\providecommand \@url [1]{\endgroup\@href {#1}{\urlprefix }}%
\providecommand \urlprefix  [0]{URL }%
\providecommand \Eprint [0]{\href }%
\providecommand \doibase [0]{https://doi.org/}%
\providecommand \selectlanguage [0]{\@gobble}%
\providecommand \bibinfo  [0]{\@secondoftwo}%
\providecommand \bibfield  [0]{\@secondoftwo}%
\providecommand \translation [1]{[#1]}%
\providecommand \BibitemOpen [0]{}%
\providecommand \bibitemStop [0]{}%
\providecommand \bibitemNoStop [0]{.\EOS\space}%
\providecommand \EOS [0]{\spacefactor3000\relax}%
\providecommand \BibitemShut  [1]{\csname bibitem#1\endcsname}%
\let\auto@bib@innerbib\@empty
\bibitem [{\citenamefont {Pang}\ \emph {et~al.}(2004)\citenamefont {Pang},
  \citenamefont {Ping}, \citenamefont {Wang}, \citenamefont {Goldman},\ and\
  \citenamefont {Zhao}}]{PangPRC2004}%
  \BibitemOpen
  \bibfield  {author} {\bibinfo {author} {\bibfnamefont {H.}~\bibnamefont
  {Pang}}, \bibinfo {author} {\bibfnamefont {J.}~\bibnamefont {Ping}}, \bibinfo
  {author} {\bibfnamefont {F.}~\bibnamefont {Wang}}, \bibinfo {author}
  {\bibfnamefont {T.}~\bibnamefont {Goldman}},\ and\ \bibinfo {author}
  {\bibfnamefont {E.}~\bibnamefont {Zhao}},\ }\bibfield  {title} {\bibinfo
  {title} {{High strangeness dibaryons in the extended quark delocalization,
  color screening model}},\ }\href {https://doi.org/10.1103/PhysRevC.69.065207}
  {\bibfield  {journal} {\bibinfo  {journal} {Phys. Rev. C}\ }\textbf {\bibinfo
  {volume} {69}},\ \bibinfo {pages} {065207} (\bibinfo {year}
  {2004})}\BibitemShut {NoStop}%
\bibitem [{\citenamefont {Huang}\ \emph {et~al.}(2015)\citenamefont {Huang},
  \citenamefont {Ping},\ and\ \citenamefont {Wang}}]{HuangPRC922015}%
  \BibitemOpen
  \bibfield  {author} {\bibinfo {author} {\bibfnamefont {H.}~\bibnamefont
  {Huang}}, \bibinfo {author} {\bibfnamefont {J.}~\bibnamefont {Ping}},\ and\
  \bibinfo {author} {\bibfnamefont {F.}~\bibnamefont {Wang}},\ }\bibfield
  {title} {\bibinfo {title} {{Further study of the
  $N\mathrm{\ensuremath{\Omega}}$ dibaryon within constituent quark models}},\
  }\href {https://doi.org/10.1103/PhysRevC.92.065202} {\bibfield  {journal}
  {\bibinfo  {journal} {Phys. Rev. C}\ }\textbf {\bibinfo {volume} {92}},\
  \bibinfo {pages} {065202} (\bibinfo {year} {2015})}\BibitemShut {NoStop}%
\bibitem [{\citenamefont {Haidenbauer}\ \emph {et~al.}(2017)\citenamefont
  {Haidenbauer}, \citenamefont {Petschauer}, \citenamefont {Kaiser},
  \citenamefont {Mei{\ss}ner},\ and\ \citenamefont {Weise}}]{Haidenbauer2017}%
  \BibitemOpen
  \bibfield  {author} {\bibinfo {author} {\bibfnamefont {J.}~\bibnamefont
  {Haidenbauer}}, \bibinfo {author} {\bibfnamefont {S.}~\bibnamefont
  {Petschauer}}, \bibinfo {author} {\bibfnamefont {N.}~\bibnamefont {Kaiser}},
  \bibinfo {author} {\bibfnamefont {U.-G.}\ \bibnamefont {Mei{\ss}ner}},\ and\
  \bibinfo {author} {\bibfnamefont {W.}~\bibnamefont {Weise}},\ }\bibfield
  {title} {\bibinfo {title} {{Scattering of decuplet baryons in chiral
  effective field theory}},\ }\href
  {https://doi.org/10.1140/epjc/s10052-017-5309-4} {\bibfield  {journal}
  {\bibinfo  {journal} {Eur. Phys. J. C}\ }\textbf {\bibinfo {volume} {77}},\
  \bibinfo {pages} {760} (\bibinfo {year} {2017})}\BibitemShut {NoStop}%
\bibitem [{\citenamefont {Sekihara}\ \emph {et~al.}(2018)\citenamefont
  {Sekihara}, \citenamefont {Kamiya},\ and\ \citenamefont
  {Hyodo}}]{Sekihara2018}%
  \BibitemOpen
  \bibfield  {author} {\bibinfo {author} {\bibfnamefont {T.}~\bibnamefont
  {Sekihara}}, \bibinfo {author} {\bibfnamefont {Y.}~\bibnamefont {Kamiya}},\
  and\ \bibinfo {author} {\bibfnamefont {T.}~\bibnamefont {Hyodo}},\ }\bibfield
   {title} {\bibinfo {title} {$\mathrm{N} \mathrm{\ensuremath{\Omega}}$
  interaction: Meson exchanges, inelastic channels, and quasibound state},\
  }\href {https://doi.org/10.1103/PhysRevC.98.015205} {\bibfield  {journal}
  {\bibinfo  {journal} {Phys. Rev. C}\ }\textbf {\bibinfo {volume} {98}},\
  \bibinfo {pages} {015205} (\bibinfo {year} {2018})}\BibitemShut {NoStop}%
\bibitem [{\citenamefont {Feng}\ \emph {et~al.}(2026)\citenamefont {Feng},
  \citenamefont {Wang},\ and\ \citenamefont {Zhao}}]{PomeronPRD2026}%
  \BibitemOpen
  \bibfield  {author} {\bibinfo {author} {\bibfnamefont {Y.-J.}\ \bibnamefont
  {Feng}}, \bibinfo {author} {\bibfnamefont {Q.}~\bibnamefont {Wang}},\ and\
  \bibinfo {author} {\bibfnamefont {Q.}~\bibnamefont {Zhao}},\ }\bibfield
  {title} {\bibinfo {title} {{Proton-Omega bound state revisited via the
  Pomeron exchange model}},\ }\href {https://doi.org/10.1103/wjc5-cvfz}
  {\bibfield  {journal} {\bibinfo  {journal} {Phys. Rev. D}\ }\textbf {\bibinfo
  {volume} {114}},\ \bibinfo {pages} {094002} (\bibinfo {year}
  {2026})}\BibitemShut {NoStop}%
\bibitem [{\citenamefont {Etminan}\ \emph {et~al.}(2014)\citenamefont {Etminan}
  \emph {et~al.}}]{etminan2014}%
  \BibitemOpen
  \bibfield  {author} {\bibinfo {author} {\bibfnamefont {F.}~\bibnamefont
  {Etminan}} \emph {et~al.},\ }\bibfield  {title} {\bibinfo {title} {Spin-2 $
  \mathrm{N} \mathrm{\ensuremath{\Omega}} $ dibaryon from lattice $
  \mathrm{QCD}$},\ }\href
  {https://doi.org/https://doi.org/10.1016/j.nuclphysa.2014.05.014} {\bibfield
  {journal} {\bibinfo  {journal} {Nucl. Phys. A}\ }\textbf {\bibinfo {volume}
  {928}},\ \bibinfo {pages} {89} (\bibinfo {year} {2014})},\ \bibinfo {note}
  {special Issue Dedicated to the Memory of Gerald E Brown
  (1926-2013)}\BibitemShut {NoStop}%
\bibitem [{\citenamefont {Iritani}\ \emph {et~al.}(2019)\citenamefont {Iritani}
  \emph {et~al.}}]{Iritani2019prb}%
  \BibitemOpen
  \bibfield  {author} {\bibinfo {author} {\bibfnamefont {T.}~\bibnamefont
  {Iritani}} \emph {et~al.},\ }\bibfield  {title} {\bibinfo {title} {$
  \mathrm{N} \mathrm{\ensuremath{\Omega}} $ dibaryon from lattice $
  \mathrm{QCD}$ near the physical point},\ }\href
  {https://doi.org/https://doi.org/10.1016/j.physletb.2019.03.050} {\bibfield
  {journal} {\bibinfo  {journal} {Phys. Lett. B}\ }\textbf {\bibinfo {volume}
  {792}},\ \bibinfo {pages} {284} (\bibinfo {year} {2019})}\BibitemShut
  {NoStop}%
\bibitem [{\citenamefont {Zhang}\ \emph {et~al.}(2026)\citenamefont {Zhang},
  \citenamefont {Lyu}, \citenamefont {Aoki}, \citenamefont {Doi}, \citenamefont
  {Hatsuda}, \citenamefont {Zhang},\ and\ \citenamefont {Ma}}]{zhang2026}%
  \BibitemOpen
  \bibfield  {author} {\bibinfo {author} {\bibfnamefont {L.}~\bibnamefont
  {Zhang}}, \bibinfo {author} {\bibfnamefont {Y.}~\bibnamefont {Lyu}}, \bibinfo
  {author} {\bibfnamefont {S.}~\bibnamefont {Aoki}}, \bibinfo {author}
  {\bibfnamefont {T.}~\bibnamefont {Doi}}, \bibinfo {author} {\bibfnamefont
  {T.}~\bibnamefont {Hatsuda}}, \bibinfo {author} {\bibfnamefont
  {S.}~\bibnamefont {Zhang}},\ and\ \bibinfo {author} {\bibfnamefont {Y.-G.}\
  \bibnamefont {Ma}},\ }\href {https://arxiv.org/abs/2611.16989} {\bibinfo
  {title} {{$N\Omega$ Dibaryon near Unitarity from Physical-Point Lattice
  QCD}}} (\bibinfo {year} {2026}),\ \Eprint {https://arxiv.org/abs/2611.16989}
  {arXiv:2611.16989 [hep-lat]} \BibitemShut {NoStop}%
\bibitem [{\citenamefont {Morita}\ \emph {et~al.}(2016)\citenamefont {Morita}
  \emph {et~al.}}]{Morita2016}%
  \BibitemOpen
  \bibfield  {author} {\bibinfo {author} {\bibfnamefont {K.}~\bibnamefont
  {Morita}} \emph {et~al.},\ }\bibfield  {title} {\bibinfo {title} {Probing
  multistrange dibaryons with proton-omega correlations in high-energy heavy
  ion collisions},\ }\href {https://doi.org/10.1103/PhysRevC.94.031901}
  {\bibfield  {journal} {\bibinfo  {journal} {Phys. Rev. C}\ }\textbf {\bibinfo
  {volume} {94}},\ \bibinfo {pages} {031901} (\bibinfo {year}
  {2016})}\BibitemShut {NoStop}%
\bibitem [{\citenamefont {Adam}\ \emph {et~al.}(2019)\citenamefont {Adam} \emph
  {et~al.}}]{STAR2019}%
  \BibitemOpen
  \bibfield  {author} {\bibinfo {author} {\bibfnamefont {J.}~\bibnamefont
  {Adam}} \emph {et~al.} (\bibinfo {collaboration} {STAR Collaboration}),\
  }\bibfield  {title} {\bibinfo {title} {The
  proton-$\mathrm{\ensuremath{\Omega}}$ correlation function in
  $\mathrm{Au}+\mathrm{Au}$ collisions at $ s\mathrm{NN}=200\text{
  }\mathrm{GeV}$},\ }\href
  {https://doi.org/https://doi.org/10.1016/j.physletb.2019.01.055} {\bibfield
  {journal} {\bibinfo  {journal} {Phys. Lett. B}\ }\textbf {\bibinfo {volume}
  {790}},\ \bibinfo {pages} {490} (\bibinfo {year} {2019})}\BibitemShut
  {NoStop}%
\bibitem [{\citenamefont {Zhang}\ \emph {et~al.}(2022)\citenamefont {Zhang},
  \citenamefont {Zhang},\ and\ \citenamefont {Ma}}]{zhang2021production}%
  \BibitemOpen
  \bibfield  {author} {\bibinfo {author} {\bibfnamefont {L.}~\bibnamefont
  {Zhang}}, \bibinfo {author} {\bibfnamefont {S.}~\bibnamefont {Zhang}},\ and\
  \bibinfo {author} {\bibfnamefont {Y.-G.}\ \bibnamefont {Ma}},\ }\bibfield
  {title} {\bibinfo {title} {{Production of $\mathrm{\ensuremath{\Omega}}NN$
  and $\mathrm{\ensuremath{\Omega \Omega}}N$ in ultra-relativistic heavy-ion
  collisions}},\ }\href {https://doi.org/10.1140/epjc/s10052-022-10336-7}
  {\bibfield  {journal} {\bibinfo  {journal} {Eur. Phys. J. C}\ }\textbf
  {\bibinfo {volume} {82}},\ \bibinfo {pages} {1} (\bibinfo {year}
  {2022})}\BibitemShut {NoStop}%
\bibitem [{\citenamefont {Etminan}(2025)}]{EtminanPRC2025-111}%
  \BibitemOpen
  \bibfield  {author} {\bibinfo {author} {\bibfnamefont {F.}~\bibnamefont
  {Etminan}},\ }\bibfield  {title} {\bibinfo {title} {Femtoscopic study of the
  $\mathrm{\ensuremath{\Omega}}\ensuremath{\alpha}$ interaction in heavy-ion
  collisions},\ }\href {https://doi.org/10.1103/PhysRevC.111.014912} {\bibfield
   {journal} {\bibinfo  {journal} {Phys. Rev. C}\ }\textbf {\bibinfo {volume}
  {111}},\ \bibinfo {pages} {014912} (\bibinfo {year} {2025})}\BibitemShut
  {NoStop}%
\bibitem [{\citenamefont {Garcilazo}\ and\ \citenamefont
  {Valcarce}(2018)}]{Garcilazo2018}%
  \BibitemOpen
  \bibfield  {author} {\bibinfo {author} {\bibfnamefont {H.}~\bibnamefont
  {Garcilazo}}\ and\ \bibinfo {author} {\bibfnamefont {A.}~\bibnamefont
  {Valcarce}},\ }\bibfield  {title} {\bibinfo {title}
  {$\mathrm{\ensuremath{\Omega}}d$ bound state},\ }\href
  {https://doi.org/10.1103/PhysRevC.98.024002} {\bibfield  {journal} {\bibinfo
  {journal} {Phys. Rev. C}\ }\textbf {\bibinfo {volume} {98}},\ \bibinfo
  {pages} {024002} (\bibinfo {year} {2018})}\BibitemShut {NoStop}%
\bibitem [{\citenamefont {Garcilazo}\ and\ \citenamefont
  {Valcarce}(2019)}]{Garcilazo2019}%
  \BibitemOpen
  \bibfield  {author} {\bibinfo {author} {\bibfnamefont {H.}~\bibnamefont
  {Garcilazo}}\ and\ \bibinfo {author} {\bibfnamefont {A.}~\bibnamefont
  {Valcarce}},\ }\bibfield  {title} {\bibinfo {title}
  {$\mathrm{\ensuremath{\Omega}}\mathit{NN}$ and
  $\mathrm{\ensuremath{\Omega}}\mathrm{\ensuremath{\Omega}}\mathrm{N}$
  states},\ }\href {https://doi.org/10.1103/PhysRevC.99.014001} {\bibfield
  {journal} {\bibinfo  {journal} {Phys. Rev. C}\ }\textbf {\bibinfo {volume}
  {99}},\ \bibinfo {pages} {014001} (\bibinfo {year} {2019})}\BibitemShut
  {NoStop}%
\bibitem [{\citenamefont {Etminan}\ \emph {et~al.}(2023)\citenamefont
  {Etminan}, \citenamefont {Sanchuli},\ and\ \citenamefont
  {Firoozabadi}}]{ETMINAN2023122639}%
  \BibitemOpen
  \bibfield  {author} {\bibinfo {author} {\bibfnamefont {F.}~\bibnamefont
  {Etminan}}, \bibinfo {author} {\bibfnamefont {Z.}~\bibnamefont {Sanchuli}},\
  and\ \bibinfo {author} {\bibfnamefont {M.~M.}\ \bibnamefont {Firoozabadi}},\
  }\bibfield  {title} {\bibinfo {title} {{Geometrical properties of $
  \ensuremath{\Omega NN} $ three-body states by realistic $ \ensuremath{NN} $
  and first principles Lattice QCD $ \ensuremath{\Omega N $} potentials}},\
  }\href {https://doi.org/https://doi.org/10.1016/j.nuclphysa.2023.122639}
  {\bibfield  {journal} {\bibinfo  {journal} {Nucl. Phys. A}\ }\textbf
  {\bibinfo {volume} {1033}},\ \bibinfo {pages} {122639} (\bibinfo {year}
  {2023})}\BibitemShut {NoStop}%
\bibitem [{\citenamefont {Filikhin}\ \emph {et~al.}(2026)\citenamefont
  {Filikhin}, \citenamefont {Kezerashvili},\ and\ \citenamefont
  {Vlahovic}}]{FilikhinPRC2025OmegaNuclei}%
  \BibitemOpen
  \bibfield  {author} {\bibinfo {author} {\bibfnamefont {I.}~\bibnamefont
  {Filikhin}}, \bibinfo {author} {\bibfnamefont {R.~Y.}\ \bibnamefont
  {Kezerashvili}},\ and\ \bibinfo {author} {\bibfnamefont {B.}~\bibnamefont
  {Vlahovic}},\ }\bibfield  {title} {\bibinfo {title} {{Bound states of
  $\mathrm{\ensuremath{\Omega}}$ baryons in light nuclei}},\ }\href
  {https://doi.org/10.1103/cjvn-7kph} {\bibfield  {journal} {\bibinfo
  {journal} {Phys. Rev. C}\ }\textbf {\bibinfo {volume} {113}},\ \bibinfo
  {pages} {015201} (\bibinfo {year} {2026})}\BibitemShut {NoStop}%
\bibitem [{\citenamefont {Malfliet}\ and\ \citenamefont
  {Tjon}(1969)}]{MALFLIET1969161}%
  \BibitemOpen
  \bibfield  {author} {\bibinfo {author} {\bibfnamefont {R.}~\bibnamefont
  {Malfliet}}\ and\ \bibinfo {author} {\bibfnamefont {J.}~\bibnamefont
  {Tjon}},\ }\bibfield  {title} {\bibinfo {title} {Solution of the
  $\mathrm{Faddeev}$ equations for the triton problem using local two-particle
  interactions},\ }\href
  {https://doi.org/https://doi.org/10.1016/0375-9474(69)90775-1} {\bibfield
  {journal} {\bibinfo  {journal} {Nucl. Phys. A}\ }\textbf {\bibinfo {volume}
  {127}},\ \bibinfo {pages} {161} (\bibinfo {year} {1969})}\BibitemShut
  {NoStop}%
\bibitem [{\citenamefont {Mohr}\ \emph {et~al.}(2025)\citenamefont {Mohr},
  \citenamefont {Newell}, \citenamefont {Taylor},\ and\ \citenamefont
  {Tiesinga}}]{RevModPhys.97.025002}%
  \BibitemOpen
  \bibfield  {author} {\bibinfo {author} {\bibfnamefont {P.~J.}\ \bibnamefont
  {Mohr}}, \bibinfo {author} {\bibfnamefont {D.~B.}\ \bibnamefont {Newell}},
  \bibinfo {author} {\bibfnamefont {B.~N.}\ \bibnamefont {Taylor}},\ and\
  \bibinfo {author} {\bibfnamefont {E.}~\bibnamefont {Tiesinga}},\ }\bibfield
  {title} {\bibinfo {title} {{CODATA recommended values of the fundamental
  physical constants: 2022}},\ }\href
  {https://doi.org/10.1103/RevModPhys.97.025002} {\bibfield  {journal}
  {\bibinfo  {journal} {Rev. Mod. Phys.}\ }\textbf {\bibinfo {volume} {97}},\
  \bibinfo {pages} {025002} (\bibinfo {year} {2025})}\BibitemShut {NoStop}%
\bibitem [{\citenamefont {Can}\ \emph {et~al.}(2015)\citenamefont {Can},
  \citenamefont {Erkol}, \citenamefont {Oka},\ and\ \citenamefont
  {Takahashi}}]{PhysRevD.92.114515}%
  \BibitemOpen
  \bibfield  {author} {\bibinfo {author} {\bibfnamefont {K.~U.}\ \bibnamefont
  {Can}}, \bibinfo {author} {\bibfnamefont {G.}~\bibnamefont {Erkol}}, \bibinfo
  {author} {\bibfnamefont {M.}~\bibnamefont {Oka}},\ and\ \bibinfo {author}
  {\bibfnamefont {T.~T.}\ \bibnamefont {Takahashi}},\ }\bibfield  {title}
  {\bibinfo {title} {{Look inside charmed-strange baryons from lattice QCD}},\
  }\href {https://doi.org/10.1103/PhysRevD.92.114515} {\bibfield  {journal}
  {\bibinfo  {journal} {Phys. Rev. D}\ }\textbf {\bibinfo {volume} {92}},\
  \bibinfo {pages} {114515} (\bibinfo {year} {2015})}\BibitemShut {NoStop}%
\bibitem [{\citenamefont {Lyu}\ \emph {et~al.}(2021)\citenamefont {Lyu},
  \citenamefont {Tong}, \citenamefont {Sugiura}, \citenamefont {Aoki},
  \citenamefont {Doi}, \citenamefont {Hatsuda}, \citenamefont {Meng},\ and\
  \citenamefont {Miyamoto}}]{LyuPRL2021}%
  \BibitemOpen
  \bibfield  {author} {\bibinfo {author} {\bibfnamefont {Y.}~\bibnamefont
  {Lyu}}, \bibinfo {author} {\bibfnamefont {H.}~\bibnamefont {Tong}}, \bibinfo
  {author} {\bibfnamefont {T.}~\bibnamefont {Sugiura}}, \bibinfo {author}
  {\bibfnamefont {S.}~\bibnamefont {Aoki}}, \bibinfo {author} {\bibfnamefont
  {T.}~\bibnamefont {Doi}}, \bibinfo {author} {\bibfnamefont {T.}~\bibnamefont
  {Hatsuda}}, \bibinfo {author} {\bibfnamefont {J.}~\bibnamefont {Meng}},\ and\
  \bibinfo {author} {\bibfnamefont {T.}~\bibnamefont {Miyamoto}},\ }\bibfield
  {title} {\bibinfo {title} {{Dibaryon with Highest Charm Number near Unitarity
  from Lattice QCD}},\ }\href {https://doi.org/10.1103/PhysRevLett.127.072003}
  {\bibfield  {journal} {\bibinfo  {journal} {Phys. Rev. Lett.}\ }\textbf
  {\bibinfo {volume} {127}},\ \bibinfo {pages} {072003} (\bibinfo {year}
  {2021})}\BibitemShut {NoStop}%
\bibitem [{\citenamefont {Hiyama}\ \emph {et~al.}(2003)\citenamefont {Hiyama},
  \citenamefont {Kino},\ and\ \citenamefont {Kamimura}}]{HIYAMA2003223}%
  \BibitemOpen
  \bibfield  {author} {\bibinfo {author} {\bibfnamefont {E.}~\bibnamefont
  {Hiyama}}, \bibinfo {author} {\bibfnamefont {Y.}~\bibnamefont {Kino}},\ and\
  \bibinfo {author} {\bibfnamefont {M.}~\bibnamefont {Kamimura}},\ }\bibfield
  {title} {\bibinfo {title} {{Gaussian expansion method for few-body
  systems}},\ }\href
  {https://doi.org/https://doi.org/10.1016/S0146-6410(03)90015-9} {\bibfield
  {journal} {\bibinfo  {journal} {Prog. Part. Nucl. Phys.}\ }\textbf {\bibinfo
  {volume} {51}},\ \bibinfo {pages} {223} (\bibinfo {year} {2003})}\BibitemShut
  {NoStop}%
\bibitem [{\citenamefont {Etminan}\ and\ \citenamefont
  {Happ}(2026)}]{etminan2026nnomegaccc}%
  \BibitemOpen
  \bibfield  {author} {\bibinfo {author} {\bibfnamefont {F.}~\bibnamefont
  {Etminan}}\ and\ \bibinfo {author} {\bibfnamefont {L.}~\bibnamefont {Happ}},\
  }\href {https://arxiv.org/abs/2508.12276} {\bibinfo {title} {{Probing
  $NN\Omega_{ccc}$ three-body systems with the modern QCD $N\Omega_{ccc}$
  interaction}}} (\bibinfo {year} {2026}),\ \Eprint
  {https://arxiv.org/abs/2508.12276} {arXiv:2508.12276 [nucl-th]} \BibitemShut
  {NoStop}%
\bibitem [{\citenamefont {Happ}(2026{\natexlab{a}})}]{Happ2026a}%
  \BibitemOpen
  \bibfield  {author} {\bibinfo {author} {\bibfnamefont {L.}~\bibnamefont
  {Happ}},\ }\bibfield  {title} {\bibinfo {title} {Fewbodytoolkit.jl: A julia
  package for solving quantum few-body problems},\ }\href
  {https://doi.org/10.21468/SciPostPhysCodeb.81} {\bibfield  {journal}
  {\bibinfo  {journal} {SciPost Phys. Codebases}\ ,\ \bibinfo {pages} {81}}
  (\bibinfo {year} {2026}{\natexlab{a}})}\BibitemShut {NoStop}%
\bibitem [{\citenamefont {Happ}(2026{\natexlab{b}})}]{Happ2026b}%
  \BibitemOpen
  \bibfield  {author} {\bibinfo {author} {\bibfnamefont {L.}~\bibnamefont
  {Happ}},\ }\bibfield  {title} {\bibinfo {title} {Codebase release 0.5 for
  fewbodytoolkit.jl},\ }\href
  {https://doi.org/10.21468/SciPostPhysCodeb.81-r0.5} {\bibfield  {journal}
  {\bibinfo  {journal} {SciPost Phys. Codebases}\ ,\ \bibinfo {pages} {81}}
  (\bibinfo {year} {2026}{\natexlab{b}})}\BibitemShut {NoStop}%
\end{thebibliography}%
\end{document}